\documentclass[twocolumn,showpacs,preprintnumbers,amsmath,amssymb]{revtex4}

\usepackage{graphicx}
\usepackage{dcolumn}
\usepackage{bm}

\begin{document}

\title{Isospin dependence of the critical quark-deconfinement densities}

\author{M. Di Toro$^1$, A. Drago$^2$, V. Greco$^1$}
\author{A. Lavagno$^3$}

\affiliation{
$^1$ Universit\`a di Catania and INFN, Lab. Nazionale
del Sud, 95123 Catania, Italy\\
$^2$ Dip. di Fisica, Universit{\`a} di Ferrara 
and INFN, Sezione di Ferrara, 44100 Ferrara, Italy\\
$^3$ Dip. di Fisica, Politecnico di Torino
and INFN, Sezione di Torino, 10129 Torino, Italy}

\begin{abstract}
We explore the dependence of the critical density, 
separating hadronic matter from a mixed phase of quarks and hadrons,
on the ratio $Z/A$. We use both the MIT bag model
and the Color Dielectric Model to describe the quark dynamics,
while for the hadronic phase we employ
various relativistic equations of state.
We find that, if the parameters of quark models are fixed
so that the existence of quark stars is allowed, then the critical density
drops dramatically in the range
$Z/A \sim $ 0.3--0.4. Moreover, for $Z/A \sim $ 0.3
the critical density is only slightly larger than the saturation density of
symmetric nuclear-matter. This opens the
possibility to verify the Witten-Bodmer hypothesis on absolute
stability of quark matter using ground-based experiments in which
neutron-rich nuclei are tested. 
\vspace{1pc}
\end{abstract}


\maketitle

Hadronic matter is expected to undergo a phase transition 
into a deconfined phase of quarks and gluons at large densities 
and/or high temperatures. On very general grounds,
the transition's critical densities are expected to depend
on the isospin of the system. 
Up to now, experimental data on the phase transition have been 
extracted from
high-energy scattering of almost isospin-symmetric nuclei, having
a proton fraction $Z/A\sim$ 0.4--0.5. The 
analysis of observations of neutron stars,
which are composed of $\beta$-stable matter for which
$Z/A\lesssim$ 0.1, can also provide hints
on the structure of extremely asymmetric matter. 
No data on the quark deconfinement transition 
is at the moment available for intermediate values of
$Z/A$. Recently it has been proposed
by several groups to produce and study unstable neutron-rich 
nuclei (for a updated review of the status
of these projects see e.g. \cite{Jonson:2002,Grunder:2002}). 
As we will show,
these new experiments open the possibility to explore in laboratory the isospin
dependence of the critical densities.

The information coming from experiments with ultra-relativistic heavy ions
is that, for symmetric or nearly symmetric nuclear matter, 
the critical energy density must be considerably larger 
than nuclear matter 
energy density at saturation.
Concerning non-symmetric matter, general arguments based on Pauli principle
suggest that the critical density decreases with $Z/A$. 
We want to study in particular the range $Z/A\sim$ 0.3--0.4, which can be
explored in radioactive nuclear beam facilities.
Here and in the
following we are mainly interested in the critical density 
separating pure hadronic matter from a mixed phase of hadrons
and quarks. The second critical density, separating
the mixed phase from the pure quark matter phase, is in general
reachable only in high energy experiments, while we are
mainly interested in intermediate energy experiments.

A sistematic study of the isospin dependence of the critical
densities has been performed up to now, to our knowledge, only by Mueller
\cite{Mueller:1997}. In that paper only one set of 
model parameters values is explored. The conclusion of \cite{Mueller:1997}
is that, moving from symmetric nuclei to nuclei having $Z/A\sim 0.3$,
the critical density is reduced by roughly 10$\%$. In this Letter we will
explore in a more sistematic way the model parameters. In particular we
will be interested in those parameters sets which 
would allow the existence of quarks stars
\cite{Alcock:1986hz,Haensel:1986}, i.e. parameters sets
for which the so-called Witten-Bodmer hypothesis is satisfied
\cite{Witten:1984rs,Bodmer:1971we}.
According to that hypothesis, a state made of an approximately equal number 
of up, down and strange quarks
can have an energy per baryon number $E/A$ smaller
than that of iron ($E_{Fe}\approx 930$ MeV).
To satisfy the Witten-Bodmer hypothesis, strong constraints
on quark model parameters have to be imposed. For instance, using
the MIT bag model, the so-called
pressure-of-the-vacuum parameter $B$ must have a very small value, 
$B^{1/4}\sim$ 140--150 MeV \cite{Farhi:1984qu}. 
Assuming the Witten-Bodmer hypothesis to be true, ordinary
nuclear matter 
would be metastable. In order not to contradict the obvious stability
of normal nuclei, quark matter made of only two flavors must not be
more stable than iron. As we shall see, slightly more strict 
boundaries on parameters' value can be imposed by requiring not
only iron, but also neutron rich nuclei like e.g. lead,
to be stable.
If the Witten-Bodmer hypothesis is satisfied,
self-bound stars entirely composed of quark matter can exist 
\cite{Alcock:1986hz,Haensel:1986,Drago:2001nq}.
Recently several analysis of observational data have emphasized 
the possible existence of compact stars having very small radii,
of the order of 9 kilometers or less 
\cite{Li:1999wt,Dey:1998rz,Pons:2001px,Drake:2002bj}.  
The most widely discussed possibility to explain the observed
mass-radius relation is based on the existence of quark stars.
It is therefore particularly interesting to envisage laboratory experiments
testing the possible signatures 
of model parameters values that would allow the existence of these
extremely compact stellar objects. What we are proposing in this
Letter is to use radioactive nuclear beams to this purpose.

It is rather unlikely, at least in the near future, that neutron
rich nuclei obtainable in radioactive nuclear beam facilities
can be accelerated at very large energies, much larger than
1 GeV per nucleon.
The scenario we would like to explore corresponds therefore to the situation
realized in experiments at moderate energy, in which the temperature
of the system is at maximum of the order of few tens MeV.
In this situation,
strange quarks cannot be produced and we need only to study the 
deconfinement transition from nucleonic matter into up and down
quark matter. 

In our analysis we have explored various hadronic and quark models.
Concerning the hadronic phase, we have used the relativistic non-linear
Walecka-type models of Glendenning-Moszkowski 
(GM1, GM2, GM3) \cite{Glendenning:1991es}. 
We have also explored the possibility of enhancing the
symmetry repulsion at high barion density introducing a coupling
to a charged scalar $\delta$-meson. As remarked in Ref.\cite{Liu:2001iz},
this is fully in agreement with the spirit of the effective
field theories, and of course with the phenomenology
of the free nucleon-nucleon interaction.
For the quark phase we have considered the MIT bag model 
at first order in the strong coupling constant $\alpha_s$ 
\cite{Farhi:1984qu} and the
Color Dielectric Model (CDM) \cite{Pirner:1992im,Birse:1990cx}. 
In the latter, quarks develop
a density dependent constituent mass through their interaction
with a scalar field representing a multi-gluon state. 
After having chosen a model for the hadronic and for the quark EOS,
the deconfinement phase transition is then described by imposing Gibbs
equilibrium conditions \cite{Glendenning:1992vb}.
This technique can be considered as an effective way
to describe the multi-quark correlations which generate
the hadronization process in a more microscopic approach.

\begin{figure}
\includegraphics[scale=0.5]{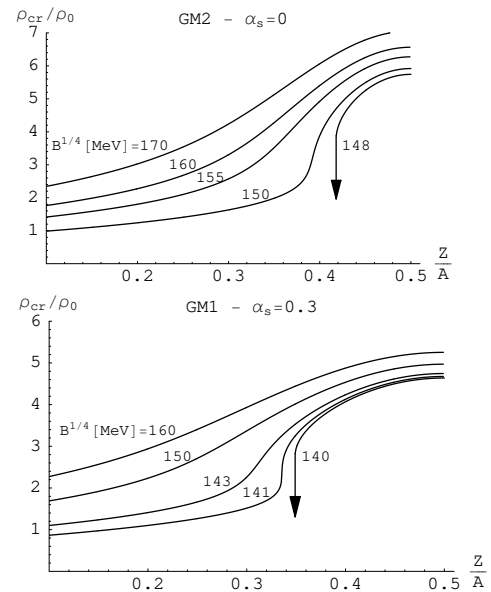}
\caption{\label{fig1}
Transition densities separating hadronic matter from mixed
quark-hadron phase. In the upper panel the GM2 parametrisation
\cite{Glendenning:1991es}
has been used for the hadronic EOS and the MIT bag model without
gluon exchange has been used for the quark EOS. In the lower panel,
GM1 parametrisation \cite{Glendenning:1991es}
for the hadronic EOS and MIT bag model with perturbative 
exchange of gluons with $\alpha_s=0.3$. The arrows
indicate that the transition density drops to very small
values and the parameter B$^{1/4}$ cannot be further reduced.
$\rho_0$ is the nuclear matter saturation density.
}
\end{figure}

In Fig. \ref{fig1} we show the
critical density $\rho_{cr}$ separating nuclear matter from
quark-nucleon mixed phase, as a function of the proton fraction
$Z/A$. The figure has been obtained using GM2 (GM1) 
parametrization for the hadronic
phase and the MIT bag model without gluons (with gluons and
$\alpha_s=0.3$) in the upper and lower
window, respectively. 
The most striking feature of the results shown in
Fig.\ref{fig1} is the
sharp decrease of $\rho_{cr}$ in the range 
$Z/A\sim$ 0.3--0.4. The lower curves
in each window correspond to parameters' values satisfying 
the Witten-Bodmer hypothesis. In the latter case, 
and for $Z/A\sim 0.3$, the critical
density is of the order of $\rho_0$. This opens the possibility to
test the deconfinement transition in low energy experiments, such
as the one performed in future RNB facilities.

\begin{figure}
\includegraphics[scale=0.5]{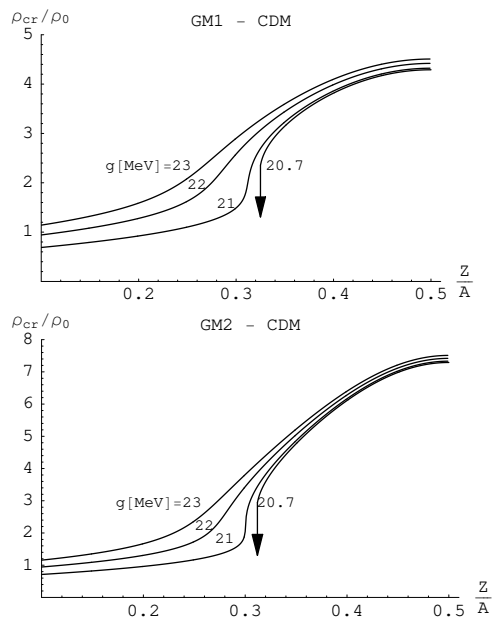}
\caption{\label{figcdm}Similar to Fig.1. The
quark EOS has been computed using the Color Dielectric Model
\cite{Drago:1995pr}. The parameter g regulates the coupling
between quarks and a scalar multigluon field, see text.}
\end{figure}

The main features of Fig.\ref{fig1} can be easely understood
if one recalls that we are investigating situations in which the minimum
of pure quark matter EOS is at an energy just above or just below
the minimum of the hadronic matter EOS. 
The first scenario is the one in which
the absolute minimum of $E/A$,
for a given value of $Z/A$, corresponds to the quark matter
EOS (this situation corresponds to very small values of the parameter $B$,
e.g. $B^{1/4}=148$ MeV in the top panel of Fig.\ref{fig1}).
In this case,
the deconfinement transition starts at very small densities, smaller
than nuclear matter saturation density.
The numerical determination of these densities is rather delicate
and we limit ourself to 
indicate with vertical arrows in Fig. \ref{fig1} the behaviour
of the critical density for a given value of $B$. If the value of
$B$ is further reduced, the vertical arrow shifts towards larger values
of $Z/A$ and therefore cannot correspond to a physically acceptable
situation, since it would imply deconfinement 
into two flavor quark matter at low densities, even for almost
symmetric nuclei. In particular we can exclude parameter's values for
which the dramatic drop in the critical density takes place for
$Z/A\ge 0.4$. In this way we define a {\it minimal} value of,
e.g., parameter $B$  of the MIT bag model. Notice that the
limit on the value of the model parameters we get in this way
is (slightly) more restrictive than the one generally adopted and based
on the stability of Fe against decay into two flavor quark matter.

The second situation is the one in which the minimum of the
quark EOS lies slightly above the hadronic minimum,
as e.g. for $B^{1/4}=150$ MeV in the top panel of Fig.\ref{fig1}. 
In this situation
the deconfinement transition starts at a density slightly smaller
than the one corresponding to the minimum of the quark EOS. 
The critical density cannot be further reduced,
since at even smaller densities the energy $E/A$
in the quark phase rises dramatically, both in the
MIT bag model and in the CDM, and therefore no mixing of 
hadronic matter with quark matter is possible at those
densities. It is also important to notice that both in the MIT
and in the CDM model, the position of the minimum of $E/A$
is near the value of nuclear matter saturation density when the energy
of the minimum is near to the one obtained from the hadronic EOS
\cite{Drago:1995pr}.

Finally, when the value of $B$ is further increased, 
the minima of the hadronic and of the quark EOS become 
more and more separated, the 
dependence of the critical density on the $Z/A$ fraction reduces
progressively, and a situation similar to the one discussed in
Ref.\cite{Mueller:1997} is reached.

\begin{figure}
\includegraphics[angle=90,scale=0.45]{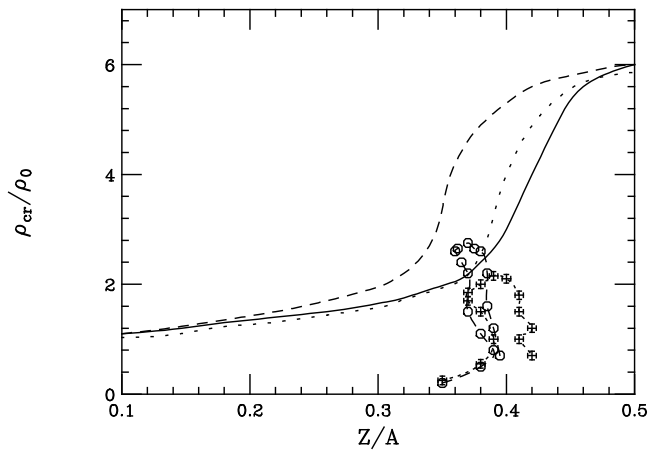}
\caption{\label{fig2}
Variation of the transition density with proton fraction for various
hadronic EOS parametrisations. Dotted line: GM2 parametrisation
\cite{Glendenning:1991es}; dashed line: NLH$\rho$ parametrisation
\cite{Liu:2001iz}; solid line: NLH($\rho + \delta$) parametrisation
\cite{Liu:2001iz}. For the quark EOS, the MIT bag model with
$B^{1/4}$=150 MeV and $\alpha_s$=0 has been used.
The points represent the path followed
in the interaction zone during a semi-central $^{132}$Sn+$^{132}$Sn
collision at 1 A.GeV (circles) and at 300 A.MeV (crosses). 
}
\end{figure}

In Fig.\ref{figcdm} the same analysis is performed using the CDM for the
quark EOS, obtaining results similar to those of
Fig.\ref{fig1}. 
In Fig.\ref{fig2} the effect of the exchange of the charged
$\delta$ meson is also considered \cite{Liu:2001iz}. 
The $\delta$-exchange potential provides an extra isospin dependence
of the EOS, and its effect shows up
in a further reduction of the
critical density in the region $0.3 \lesssim Z/A \lesssim 0.4$.
We would like to remark that the NLH$\rho$ parametrisation gives
results very similar to those obtained using the GM3 parameters set.

Let us now comment on the physical relevancy of the dramatic
reduction of the critical density in neutron rich nuclei.
Since in these nuclei neutrons presumably occupy an extended area
around the core of the nucleus (neutron skin), the density of the
very neutron-rich part can be considerably smaller than $\rho_0$.
We cannot therefore expect to find a direct
signal of deconfinement in the structure
of these nuclei, but we can look for precursor signals.
In particular, we can expect that the formation of clusters
containing six or nine quarks will be enhanced due to the reduction
of the critical deconfinement density. 
This enhancement can in turn be intepreted as a modification of 
single-nucleon properties due to the nuclear medium.
The experimental search of effects like the one we are discussing
here has a very long story, which includes the
discovery of the EMC effect~\cite{Aubert:1983xm},
that is the non trivial difference between free-nucleon
and nuclear structure functions, for which
many interpretations have been proposed 
(for a review see \cite{Arneodo:1994wf}).
In particular, models for the EMC effect
invoking the formation of multi-quark clusters have been 
\cite{Jaffe:1983rr,Carlson:1983fs} and are still quite
popular \cite{Barshay:2000dx}. 
The decrease of the deconfinement critical density obtained in our analysis
suggests a dependence of the EMC effect on the isospin,
since the probability of forming virtual multiquark bags is 
enhanced in neutron rich nuclei. This dependence
would add to the non-isoscalarity effect originated by
the different structure functions of protons
and neutrons, generally considered in the 
analyses \cite{Eskola:1998df}.
A full account of non-isoscalarity corrections
arizing from genuine nuclear physics effects has been
attempted in \cite{Barone:1999yv,Barone:2000mf}, and seems to be necessary
in the light of the preliminary analysis of CHORUS data
on deep inelastic scattering off a Pb target \cite{Oldeman:2000}.

\begin{figure}
\includegraphics[angle=90,scale=0.4,trim=0 0 -20 0]{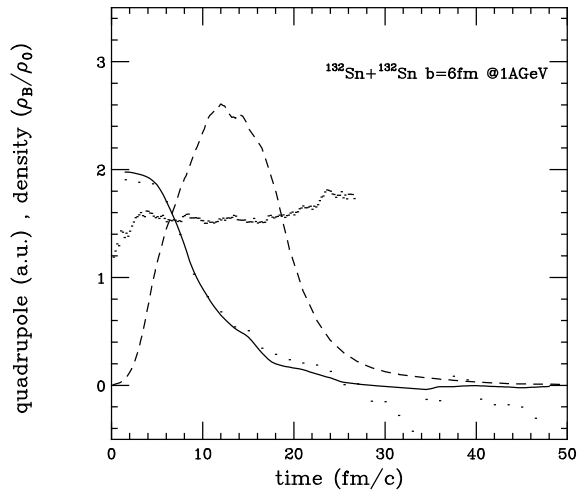}
\caption{\label{fig3}
Time evolution of the quadrupolar momentum in momenta space
(solid line) and of the density (dashed line). The simulation
examines the after-scattering
thermalisation inside a cubic cell 2.5 fm wide, located in the center
of mass of the system.  
}
\end{figure}

A more direct way to explore the reduction of the 
deconfinement critical density would be
to test the EOS of matter via scattering
of two neutron rich nuclei.
This possibility is based on the analysis
of intermediate-energy heavy-ion collisions, as 
discussed in 
Refs.\cite{Li:1998px,Li:1998ze,Li:2002qx,Li:2002yd,Baran:2001pz},
and references therein. 

To check the practical feasibility of such an experiment,
we have performed some simulations of the $^{132}$Sn + $^{132}$Sn
collision (average $Z/A$=0.38) at various energies, for
semicentral impact parameter, $b$=6 fm, just to optimize the
neutron skin effect in order to get a large asymmetry in the
interaction zone. We have used a 
Relativistic Transport Code \cite{Fuchs:1995fa},
adoptining the same effective interaction \cite{Liu:2001iz} 
of the $NLH(\rho+\delta)$
EOS of Fig.\ref{fig2} to compute the critical deconfinement density
(solid line). 
In Fig.\ref{fig2} the paths in the $(\rho,Z/A)$ plane followed
in the c.m. region during the collision are reported at energies
of 300 A.MeV (crosses) and 1 A.GeV (circles). We see that already
at 300 A.MeV we are reaching the border of the mixed phase, and we are
well inside it at 1 A.GeV. 
In order to be sure that we are really testing the Nuclear Matter EOS,
which is an equilibrium property, we have performed a check of the 
local thermalization in correspondence of the high baryon density
regions reached during the collision.
In Fig.\ref{fig3} we show that indeed, when the
maximum density is reached ($\rho\sim 2.6 \rho_0$) the quadrupole
momentum of the nucleon momentum distribution has dropped
to $\sim 10\%$ of its initial value, a signal that the system
has indeed thermalized.

The use of an harder hadronic EOS for
symmetric matter at high density would correspond to a even more
favourable situation than the one
presented in Fig.\ref{fig2}, as shown in the examples of Fig.\ref{fig1}.
Moreover the use of neutron-richer nuclei would allow to test
the EOS at smaller values of $Z/A$. To this purpose, the most promising
nuclei are the ones near the r-process path, in particular
for neutron numbers near the magic values N=82 or 126.
In these regions, the proton fraction is as low as 0.32--0.33
and these nuclei could be studied in future experiments with neutron-rich
beams \cite{Rehm:1999}.

In conclusion, our analysis supports the possibility of observing
precursor signals of the transition to a mixed quark-hadron phase
in the collision, central or semi-central, of
exotic (radioactive) heavy ions in the energy range of a few hundred
MeV per nucleon. A possible signature could be revealed through
an earlier softening of the hadronic EOS for large isospin
asymmetries, and it would be  
observed e.g. in the behaviour of the collective flows for particles
having large transverse momentum.

\medskip
\begin{acknowledgments}
It is a pleasure to thank V. Barone for many valuable suggestions.
\end{acknowledgments}



\end{document}